\documentclass[STIX2COL]{WileyNJD-v2}

\usepackage{amssymb} 
\usepackage{amsthm,amsmath}
\usepackage{multirow}
\usepackage{diagbox}
\usepackage{ragged2e}
\usepackage{graphbox}
\usepackage{hyperref}
\usepackage{tabularx}
\usepackage{makecell}

\articletype{Original Rsearch}

\begin{document}

\title{Feature graph construction with static features for malware detection}

\author[1,2]{Binghui Zou}

\author[1,2]{Chunjie Cao}

\author[1,2]{Longjuan Wang*}

\author[1,2]{Yinan Cheng}

\author[1,2]{Chenxi Dang}

\author[1,2]{Ying Liu}

\author[1,2]{Jingzhang Sun}

\authormark{Binghui Zou \textsc{et al.}}

\address[1]{\orgdiv{School of Cyberspace Security}, \orgname{Hainan University}, \orgaddress{\state{Haikou}, \country{China}}}

\address[2]{\orgdiv{Key Laboratory of Internet Information Retrieval of Hainan Province}, \orgname{Hainan University}, \orgaddress{\state{Haikou}, \country{China}}}

\corres{Longjuan Wang, School of Cyberspace Security, Hainan University, Haikou 570228, China.\\ \email{wanglongjuan@hainanu.edu.cn}}

\fundingInfo{National Natural Science Foundation of China Enterprise Innovation and Development Joint Fund (No.U19B2044), the National Key Research and Development Program of China (No.2021YFB2700600), and the Natural Science Foundation of Hainan Province (No.621MS017).}

\abstract[Abstract]{Malware can greatly compromise the integrity and trustworthiness of information and is in a constant state of evolution. Existing feature fusion-based detection methods generally overlook the correlation between features. And mere concatenation of features will reduce the model's characterization ability, lead to low detection accuracy. Moreover, these methods are susceptible to concept drift and significant degradation of the model. To address those challenges, we introduce a feature graph-based malware detection method, MFGraph, to characterize applications by learning feature-to-feature relationships to achieve improved detection accuracy while mitigating the impact of concept drift. In MFGraph, we construct a feature graph using static features extracted from binary PE files, then apply a deep graph convolutional network to learn the representation of the feature graph. Finally, we employ the representation vectors obtained from the output of a three-layer perceptron to differentiate between benign and malicious software. We evaluated our method on the EMBER dataset, and the experimental results demonstrate that it achieves an AUC score of 0.98756 on the malware detection task, outperforming other baseline models. Furthermore, the AUC score of MFGraph decreases by only 5.884\% in one year, indicating that it is the least affected by concept drift.}

\keywords{malware detection, feature graph, graph representation, concept drift}


\maketitle


\section{Introduction}\label{Introduction}

The Internet's rapid development has led to numerous network security concerns, with malware being one of the most significant threats in cyberspace. Malware has the potential to exfiltrate user data and compromise privacy, thereby inflicting substantial harm on individuals, businesses, and governments.  According to the 2020 Kaspersky Security Bulletin, approximately 10.18\% of users with networked computers subjected at least one attack of malware during the preceding year. This statistic is accompanied by the alarming discovery of an average of 360,000 new malicious files detected daily, and a striking 21 ransomware families identified within the course of 2020 alone. The situation is further compounded by attackers who are motivated by financial gain and are actively diversifying their attack channels while simultaneously employing increasingly sophisticated techniques to evade detection. A research report by DeepInstinct \cite{Cyber2021Deep}, a cybersecurity company, reveals that in 2020, there were hundreds of millions of cyberattack attempts on a daily basis. Furthermore, the report highlights a 358\% overall increase in the number of malware programs compared to 2019, with ransomware alone increasing by 435\%. According to the X-Force Threat Intelligence Index 2021 \cite{XForce2021IBM}, ransomware continues to pose the most significant threat to large organizations and enterprises. Conservative estimates suggest that ransomware authors earned profits of at least \$123 million and stole approximately 21.6 TB of data in 2020. Researching effective and consistently detectable malware models is an urgent task in cybersecurity.

Traditional malware detection approaches focus on individual features of malware, such as semantic features obtained through static methods \cite{jang2011bitshred} or behavioral features obtained through dynamic approaches \cite{Wang2018Detecting}. Dynamic analysis necessitates the simulation of the actual environment in which the application operates, which can be arduous and time-consuming. Conversely, static analysis renders a simpler and more efficient approach due to its lack of requirement for program execution. Whether employing static or dynamic analysis, the processing and selection of features can pose significant challenges when confronted with a multitude of features \cite{Babaagba2019Study,wu2023droidrl}. The quantity and significance of features can crucially impact detection accuracy, emphasizing the importance of considering both factors when developing detection methods. During feature selection, commonly utilized features include API calls, N-grams, DLLs, and opcodes, among others. While a carefully chosen combination of features may result in high detection efficiency, feature selection is not a straightforward task. While reducing the number of features can expedite the learning process and enhance detection efficiency, an insufficient number of features can lead to a decrease in detection accuracy. Furthermore, disregarding the interdependence among features and resorting to a simplistic concatenation of features can result in a significant loss of valuable information, thereby impeding the identification of potential relationships among features.

Recent research has shown that converting applications into grayscale or color images is a promising method for malware detection \cite{sun2018deep,yuan2020byte}. With the powerful feature extraction capability of Convolutional Neural Networks (CNNs), subtle differences in texture features on malicious images can be well identified, enabling malware detection and fine-grained classification between families. Although this approach does not require feature engineering, the number of bytes contained between benign software and different families of malware varies widely, which makes the difference in the size of the generated malicious images too large. Even though they can be resized to a uniform size, the images lose detailed information during the transformation process, which makes the detection and classification results unreliable. In addition, malware is constantly evolving, and existing models are often unable to achieve accurate classification in the face of unknown malware and are susceptible to conceptual drift, leading to model degradation. 

Concept drift refers to the unpredictable changes over time in the correspondence between input data and the target variable \cite{lu2018learning}. Studies have shown \cite{Ma2021Comprehensive,barbero2022transcending} that malware detection methods are susceptible to concept drift, resulting in a decline in performance, such as lower detection accuracy and F1 score values. Despite the existence of various approaches to address conceptual drift in other domains \cite{Brzezinski2014Reacting,Minku2012DDD}, the majority of these methods rely on static data sources. Given the rapid proliferation of malware variants, these techniques are not well-suited for malware detection tasks. In malware detection tasks, existing approaches tackle the concept drift problem by periodically training the model \cite{mariconti2016mamadroid} or through integration methods \cite{hu2017concept}. However, the cost of retraining is too high for the former, and the latter faces the challenge of difficult feature selection.

The emergence of graph neural network (GNN) has garnered significant research interest in recent years \cite{deng2021graph}, owing to its capacity for analyzing and characterizing graph-structured data. GNNs have demonstrated exceptional success in diverse domains, including but not limited to recommendation systems \cite{wu2022graph}, knowledge graphs \cite{wang2017knowledge}, and molecular structures \cite{you2018graph}. GNN can define learnable composite functions directly on the graph, thereby extending classical networks (e.g., long short-term memory network, CNNs, and multilayer perceptron) to more irregular and general domains. While researchers have already utilized graph neural networks to malware detection \cite{Hou2017HinDroid,Ling2022MalGraph} or classification tasks \cite{Yan2019Classifying,Sun2022Leveraging} with promising results, they have yet to consider the impact of concept drift.

To address these limitations, this research proposes MFGraph, a feature graph-based method for malware detection. First, the static features of binary PE files are extracted using LIEF \cite{LIEF2021Quarkslab}. Next, a graph structure is introduced to fuse the static features, which is modeled as a graph network to capture the relationships between different features. To learn robust representations of binary files from the constructed feature graphs, the graph embeddings are obtained by employing a deep graph convolutional network \cite{zhang2018end}, and the detection results are outputted by the classifier. Experiments conducted on the EMBER dataset containing 800K samples show that MFGraph outperforms the method of directly concatenating features in detection accuracy and can effectively reduce the effect of concept drift.

The main contributions of this study are as follows:

\begin{itemize}
	\item We propose a static feature fusion approach based on graph representation learning to construct a network of relationships between different features. The feature graph, composed of feature nodes, can provide a more comprehensive characterization of binary PE files, which can be utilized for malware detection tasks.
	
	\item We demonstrate a robust feature extraction method that utilizes deep graph convolutional networks to get the constructed feature graphs and obtain distinguishing features between benign and malicious software.
	
	\item We evaluate our proposed approach on the EMBER dataset, and the results demonstrate that MFGraph can learn the potential relationships between features and achieve the best detection accuracy compared to the baseline method of feature concatenation. Additionally, MFGraph exhibits the most stable performance and effectively mitigates the impact of concept drift.
\end{itemize}

\begin{table*}[htbp]\centering
	\caption{Comparison of proposed work with existing malware detection methods.}
	\label{ComparisonwithOthersMethods}
	\begin{tabularx}{\linewidth}{cXXXX}
		\hline
		\textbf{Works}                                                                               & \textbf{Features}                                                                       & \textbf{Models}                                                    							  & \textbf{Datasets}                                                                   						  & \textbf{Motivation}                                                                                                              \\
		\hline
		Alazab et al. \cite{Alazab2020Intelligent}                                         & Permission and API calls.                                             & ML models RF, RT, k-NN and NB            & Large datasets with 27,891 mobile apps.                                    & Combining permission requests and API calls to enhance detection model reliability.                          \\
		Wadkar et al. \cite{Wadkar2020Detecting}										& 55 features extracted from PE file.								& SVM				& Consisting of 26,245 Windows PE files belonging to 13 families. 				& Using automated and quantifiable ML techniques to detect evolutionary changes in malware families.  \\
		CruParamer \cite{chen2022cruparamer}  					& API   calls with parameter-augmented.                      & DNNs                                                     								  & 20K   execution traces of Windows PE files.                                & Using api sequences with parameters to improve   malware detection accuracy.                                              \\
		IMCEC \cite{Danish2020Image}                                  & Gray image.                                                                   & Deeper CNNs                                              						 & Malimg and unpacked   malware dataset.                                    & Reducing recognition time while improving   classification accuracy                                                       \\
		IMCFN \cite{Danish2020IMCFN}                                 & Color image.                                                                  & Fine-tuned CNNs                                          					  & Malimg and iot-android mobile dataset.                                       & Reducing computational overhead and improving classification accuracy through data enhancement.                            \\
		Scorpion \cite{Fan2018Gotcha}                                   & Heterogeneous   information network with DLL, archive, API, machine, file, and their semantic relationships. & Metagraph2vec+SVM                                       & A   large PE files dataset from Comodo Cloud Security Center.              & Modeling different entities and relationships   using heterogeneous information networks to improve detection accuracy. \\
		HomDroid \cite{Wu2021HomDroid}                            & Function  call graphs.                                                  & ML models LR, DT, 1NN, SVM, RF and 3NN & 8,198   Android apps including 3,358 covert malware and 4,840 benign apps. & Detecting  Android covert malware.                       \\
		MFGraph(ours)                                                 & Feature   Graph with nine features extracted by LIEF parser.                  & DGCNN+MLP                                      & 800K samples in ember2018.                                        & Enhancing the capability of malicious software characterization, improving detection precision and alleviating the impact of concept drift are essential. \\
		\hline
	\end{tabularx}
\end{table*}

The remainder of this paper is structured as follows: In Section \ref{Related works}, we provide a review of the relevant literature. Section \ref{Proposed method} introduces the MFGraph method and presents its details. In Section \ref{Experiments}, we describe the experimental results and systematically evaluate the performance of MFGraph. We discuss the forms of feature graph, complexity and interpretability in Section \ref{Discussion}. Finally, we present concluding remark and future work in Section \ref{Conclusion and future work}.

\section{Related work}
\label{Related works}

\subsection{Sequence-based detection methods}

Malware causes serious damage and threat to the security of computer devices in the Internet, which has caused widespread concern among anti-virus researchers. Traditional malware detection methods, such as signature-based \cite{david2017structural} or behavior-based \cite{bernardi2019dynamic}, are inadequate in detecting malicious programs. The prerequisite for achieving a high detection accuracy is the presence of such signatures or behaviors in malware training samples, and attackers can easily evade detection through obfuscation techniques. To detect unknown malware, several signature generation algorithms have been proposed. For instance, DeepSign \cite{David2015DeepSign} is an automated malware signature generation and classification method, AndroSimilar \cite{Faruki2013AndroSimilar} generates signatures to detect malicious programs by extracting statistically unlikely features, and MalGene \cite{Kirat2015MalGene} uses algorithms borrowed from bioinformatics to automatically locate system evasive behavior in call sequences, an automated technique for extracting and analyzing evasive features. However, these methods employ a relatively individual set of features that do not accurately characterize malware. Raff et al. \cite{raff2018malware} employed raw byte sequences of malware as input to a neural network, transforming the detection of raw bytes into a sequence problem that deals with over 2 million time steps. Nevertheless, the excessively long sequences come at the cost of a vast number of parameters and extreme difficulty in training.

With the advancement of Machine Learning (ML) techniques, systems utilizing ML for malware detection \cite{Hou2016Deep4MalDroid,ye2008intelligent} have become increasingly prevalent. Many researchers are currently focusing on API-based dynamic feature extraction techniques \cite{ki2015novel,pektacs2018malware}. Rabadi et al. \cite{Rabadi2020Advanced} proposed a lightweight dynamic feature extraction technique that combines API parameters and utilizes it for malware detection and classification. However, dynamic analysis is time-consuming and requires extensive feature engineering \cite{zhang2020dynamic}, whereas malware detection necessitates quick detection results. To investigate the impact of concept drift, Ma et al. \cite{Ma2021Comprehensive} compared disassembly-based, binary-based, and image-based malware detection methods and found that these methods are susceptible to concept drift, as evidenced by a decrease in F1 score scores. Additionally, Alsulami et al. \cite{Alsulami2017Lightweight} investigated the loss of malware detection performance under the influence of concept drift. Hu et al. \cite{hu2017concept} proposed an integrated learning method ENBCS to handle the concept drift in the task of Android malware detection and achieved favorable outcomes, provided that the program's permissions, operations, and code are obtained. However, accessing the code layer's functions can be challenging.

\subsection{Image-based detection methods}

CNNs have demonstrated powerful classification performance in malware detection task, which has sparked the attention of many security researchers. Nataraj et al. \cite{nataraj2011malware} first proposed to transform malware into grayscale images and classify them using ML-based methods, providing a new idea in the field of malware detection. Since then, numerous techniques for detecting image-based malware have been suggested, such as HIT4Mal\cite{vu2020hit4mal}, SERLA\cite{Jian2021novel}, DeepVisDroid\cite{bakour2021deepvisdroid}, MCFT-CNN \cite{kumar2021mcft}.  HIT4Mal is a noteworthy color coding design for malware images that leverages a combination of entropy coding and character class schemes on a Hilbert curve to achieve an accuracy of 93.01\%. SERLA is a deep neural network architecture that integrates SEResNet50, Bi-LSTM, and Attention mechanisms, and transforms malware into a three-channel malicious image. DeepVisDroid is another hybrid deep learning model for grayscale malware image classification, while MCFT-CNN accomplishes the task of classifying malicious images through transfer learning. In addition, Vasan et al. \cite{Danish2020IMCFN} transformed binary files into color images and detected malware variants by a fine-tuned CNN structure. In the same year, their other work IMCEC used grayscale images as input and learned classification using a deep CNN network. Although image-based detection methods have had some success, however, these methods overlook the potential model degradation caused by concept drift that may occur within real-world environments.

\subsection{Graph structure-based detection methods}

The primary objective of GNN is to acquire node-level or graph-level representations of input graphs, consequently offering a more generalized approach for data abstraction. HomDroid \cite{Wu2021HomDroid} detects covert malware by assessing the homogeneity of call graphs, exhibiting better detection accuracy than PerDroid, Drebin, and MaMaDroid. IntDroid \cite{Zou2021IntDroid} regards an application's function call graph as a complex social network and represents the graph's semantic features by computing the average closeness between sensitive API calls and the central node. To achieve Android malware detection, Hou et al. \cite{Hou2017HinDroid} implemented the first structured heterogeneous information networks. Fan et al. \cite{Fan2018Gotcha} developed the malware detection system Scorpion. However, since it is a node classification problem, the graph structure needs to be reorganized for node embedding when new nodes arrive. Yan et al. \cite{Yan2019Classifying} utilized graph convolutional networks for malware classification, characterizing malware using control flow graphs and embedding structural information in control flow graphs using deep graph convolutional networks to achieve effective and efficient malware classification. However, this approach requires expensive program analysis, and they did not consider the impact of concept drift.

\begin{figure}[t]
	\centering
	\includegraphics[width=0.49\textwidth]{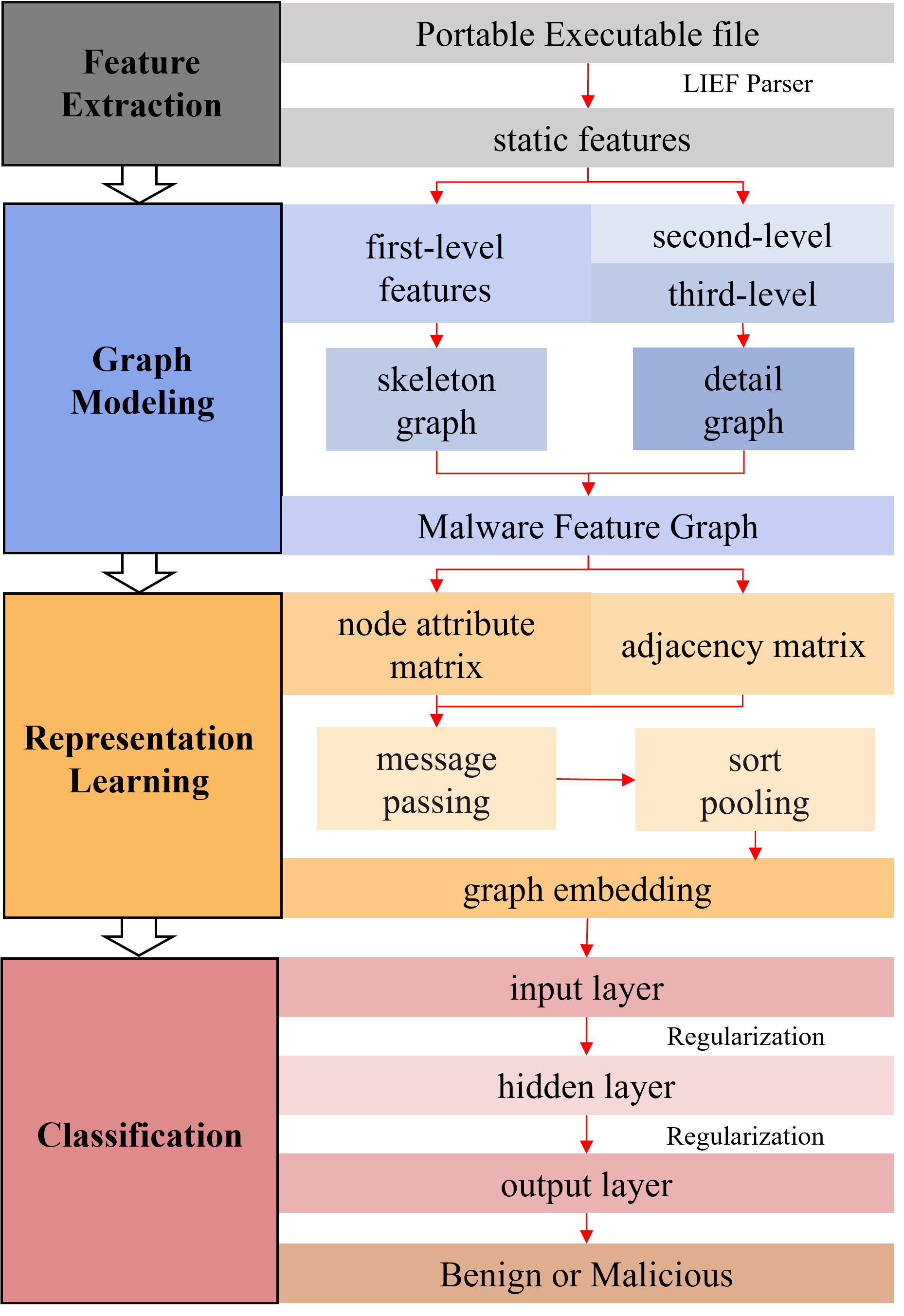}
	\caption{Workflow diagram for malware detection using static features to construct feature graphs.}\label{workflow}
\end{figure}

Compared to the previous work outlined in Table \ref{ComparisonwithOthersMethods}, the study presented in this paper differs in three primary aspects. 1) We directly extract static features of binary PE files using LIEF without complex feature selection. 2) We employ a graph structure to fuse static features to explore the potential relationships between features. 3) We examine the impact of concept drift on existing detection methods over an extended timeline.

\section{Proposed method}
\label{Proposed method}

\begin{figure}[htbp]
\begin{center}
	\label{algo-mfgraph}
	\resizebox{\linewidth}{!}{
		\begin{tabular}{l}
			\hline
			\textbf{Algorithm 1:} Feature graph construction with static features  \\
			\hline
			1: GRAPH CONSTRUCT $\left(data\right)$ \\
			2: \hspace{1cm} Open JSON file containing  PE features data: \\
			3: \hspace{1cm} Initialize empty graph $\mathbf{G}$ \\
			4: \hspace{1cm} for each section $s$ in PE features data \\
			5: \hspace{2cm} Extract features from $s$ \\
			6: \hspace{2cm} Clean and preprocess features \\
			7: \hspace{2cm} for each feature $f$ in section $s$: \\
			8: \hspace{3cm} if $f$ contains unwanted characters: \\
			9: \hspace{4cm} Remove unwanted characters from $f$ \\
			10: \hspace{3cm} if $f$ is a list: \\
			11: \hspace{4cm} Convert $f$ to indices \\
			12: \hspace{3cm} Pad $f$ with zeros to match histogram length \\
			13: \hspace{2cm} Add $s$ and its processed features to $\mathbf{G}$ \\
			14: \hspace{1cm} for each node $n$ in $\mathbf{G}$: \\
			15: \hspace{2cm} Determine connections based on feature similarity \\
			16: \hspace{2cm} Add edges to $\mathbf{G}$ based on connections \\
			17: \hspace{1cm} Write $\mathbf{G}$ to output file \\
			18: \hspace{1cm} return $\mathbf{G}$ \\
			\hline
		\end{tabular}
	}
\end{center}
\end{figure}

\begin{figure*}[htbp]
	\centering
	\includegraphics[width=0.99\textwidth]{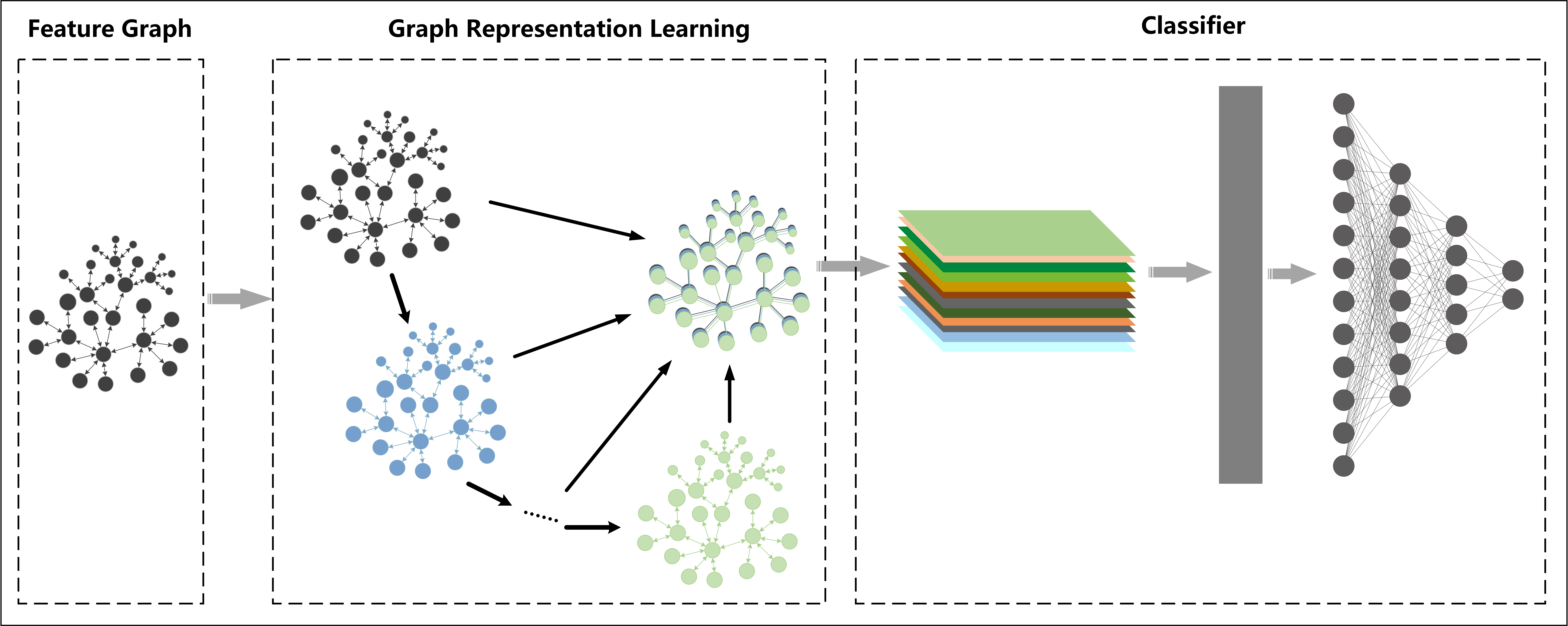}
	\caption{Process of MFGraph implementation.}\label{MFGraph}
\end{figure*}

The workflow diagram of the malware detection method we propose is shown in Figure \ref{workflow}, and the Malware Feature Graph (MFGraph) model for malware detection comprises three essential modules: Feature Graph Modeling (FGM), Graph Representation Learning (GRL), and Classifier Module (CM). As illustrated in Figure \ref{MFGraph}, FGM employs LIEF \cite{LIEF2021Quarkslab} to extract static features from the raw binary PE file, models the program as a feature graph, and utilizes a graph structure to fuse multiple features to enhance nodes with structural features and propagation characteristics of the graph. This approach captures feature-to-feature correlations for a more comprehensive characterization of the application. Subsequently, we transform malware detection into a graph binary classification problem by utilizing a deep graph convolutional network for graph representation learning and a multilayer perceptron for classification learning. 

Given a program $P$, nine static features are obtained by the LIEF parser: General, Header, Imported, Exported, Section, Byte Histogram, Byte Entropy Histogram \cite{Saxe2015Deep}, Data directories, and String. Further details of these features are presented in Table \ref{StaticDetailsLIEF}. FGM utilizes these nine static features to construct a graph G for program $P$ and the core idea of constructing the feature graph is shown in Algorithm 1. In the case of obtaining $G=\left(V,E\right)$, GRL generates the corresponding graph embedding $E_G$ through deep graph convolution and outputs the classification result $P_{out}$ via the CM, thereby enabling malware detection. During the malicious detection phase, the classifier outputs either 0 or 1, indicating benign and malicious, respectively.

\begin{table*}[htbp]\centering
	\caption{Static features and details extracted from LIEF parser.}
	\label{StaticDetailsLIEF}
	\begin{tabularx}{\linewidth}{c|X}
		\hline
		\textbf{Features}                 & \textbf{Description}                                                                                                                                                        \\
		\hline
		General                  & Includes features are: file raw size,   file virtual size, number of exported and imported functions, presence of debugging, thread local storage, resource table, relocation table, number of   signatures and symbols.                                                                                                           \\
		Header                   & Header information for PE files is contained in both COFF headers and optional headers. COFF headers hold the image file, target machine, and timestamp, while optional headers contain details on the target subsystem, image versions, commit size, file identification, DLL and linker version.
		
		\\
		Imported                 & All dynamic link libraries used by   the PE file, i.e. imported functions.                                                                                                                                                                                                                                                           \\
		Exported                 & Export symbol address, name and   serial number, most exe files have no export table, most dll files have   export table.                                                                                                                                                                                                            \\
		Section                  & Records information about each section, including name, virtual size, size, entropy, and list of strings.                          \\
		Data Directories         & Data directory table including export   table, import table, resource directory, exception directory, security   directory, relocation table, debug directory, copyright information, pointer   directory, TLS directory, load configuration directory, bind input directory,   import address table, delayed load, COM information. \\
		Byte Histogram           & Indicates the number of each byte value (256 integer values) in the PE file.                                                                                                                                                                                                                                                       \\
		Byte   Entropy Histogram & Fixed window length entropy, statistical sliding window (bytes, entropy value) pairs, in this chapter the   window length is 1024 bytes, step size is 256 bytes.               \\
		String                   & Simple statistics on strings,   including histogram, character entropy, average length, number of strings, and paths, URL links, registry  keys, short strings MZ, etc. \\
		\hline                                                                                                           
	\end{tabularx}
\end{table*}

\subsection{Feature graph construction}
\label{mfgraph_con}
\begin{figure}[t]
	\centering
	\includegraphics[width=0.39\textwidth]{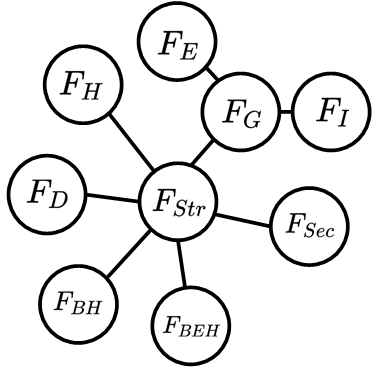}
	\caption{Feature graph constructed using nine static features: General ($F_G$), Header ($F_H$), Imported ($F_I$), Exported ($F_E$), Section ($F_{Sec}$), Byte Histogram ($F_{BH}$), Byte-Entropy Histogram ($F_{BEH}$), String ($F_{Str}$), and Data Directories ($F_D$).}
	\label{Graph_nine}
\end{figure}

To construct an appropriate feature graph, we conducted a comprehensive analysis of nine static features and integrated numerous empirical experimental results to determine the final node connections illustrated in Figure \ref{Graph_nine}, also known as the skeleton graph. We began by selecting String ($F_{Str}$) as the foundation for the feature graph due to its inclusion of global statistical data on strings found in executable files, including string count, average length, and occurrences of specific characters. Byte Histogram ($F_{BH}$) and Byte Entropy Histogram ($F_{BEH}$) offer insights into overall distribution and randomness respectively, complementing $F_{Str}$ in a mutually reinforcing manner. Additionally, section ($F_{Sec}$) and data directories ($F_D$) provide details on code, data content, and locations of various data structures within files, these are specific string attributes directly linked to $F_{Str}$. Header ($F_H$) contains metadata pertaining to PE files, while it does not contain target strings itself, it offers crucial file structure information. Moreover, general properties ($F_G$) encompass fundamental characteristics of PE files such as size, type, import and export function count etc., closely associated with $F_{Str}$.

Using the methods mentioned above, we have connected nodes for seven features, leaving only imported ($F_I$) and exported ($F_E$) features pending. The import table in $F_I$ lists functions that PE files need to call from external dynamic link libraries (DLLs), while the export table in $F_E$ lists functions provided by PE files for other programs or DLLs to use. These will be transformed into specific string features. However, they are closely linked to $F_G$ as it contains the count of imported and exported functions. Connecting $F_I$ and $F_E$ with $F_{Str}$ would lead to information redundancy and reduce the depth of the feature graph, which is not ideal for node propagation. So far, we have successfully created a node graph for nine static features.

\subsection{Feature graph modeling}

In malware detection task, it is often necessary to extract multiple features of an unknown program, followed by feature selection or feature fusion. Selecting valuable features becomes difficult with a large number of features. When processing features, a common method is feature fusion, which combines existing features or features from different levels, but simple concatenating of features does not handle the relationship between features well. Recently, feature fusion by graph representation has shown a very promising technique for feature fusion \cite{Li2019Adaptive,holzinger2021towards}, where various features are modeled as graph networks through graph structures, thus eliminating the need for tedious feature selection.

\begin{figure}[t]
	\centering
	\includegraphics[width=0.49\textwidth]{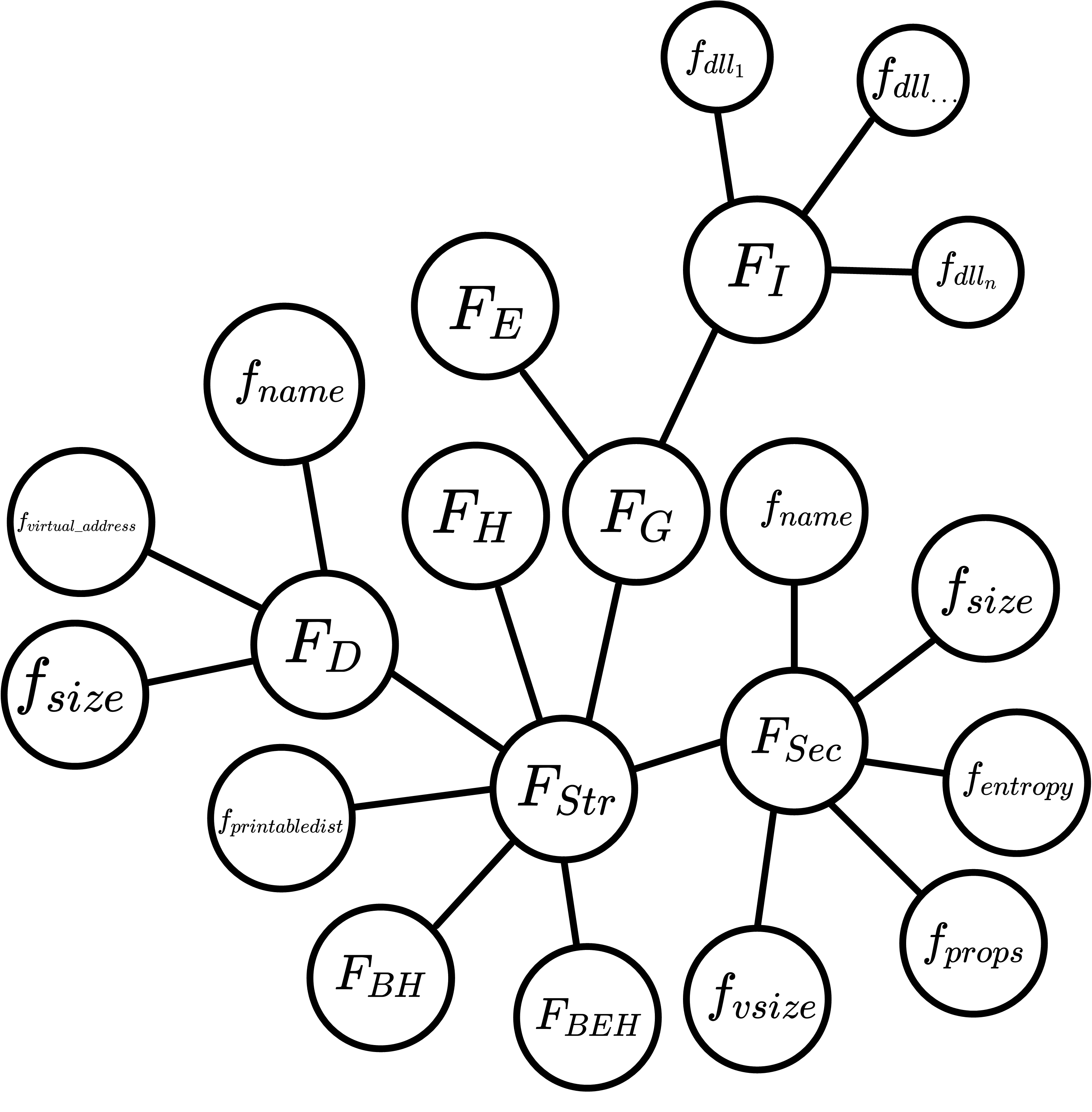}
	\caption{Feature graph modeling.}\label{feature-graph}
\end{figure}

Following the idea of feature graph modeling, we form the extracted static features \cite{anderson2018ember} into a graph network, as shown in the Figure \ref{feature-graph}, in which the edges represent the relationship between features, and the nodes represent the features of the program, which we call "feature nodes". Formally, each binary PE file can be represented as a graph $G=\left(V,E\right)$, where V denotes a set of nodes divided into major nodes $\left(Majornodes,V_{major}\right)$ and child nodes $\left(Childnodes,V_{child}\right)$. Each major node $V_{major}$ represents a static feature of a binary file. These features include: General ($F_G$), Header ($F_H$), Imported ($F_I$), Exported ($F_E$), Section ($F_{Sec}$), Byte Histogram ($F_{BH}$), Byte Entropy Histogram ($F_{BEH}$), Data directories ($F_D$), and String ($F_{Str}$). The child node $V_{child}$ is a child feature of each $V_{major}$. $F_{I}$ denotes feature Imported, $f_i$ denotes the name of the program import function, and an application importing multiple functions is denoted as $F_E=\{f_1,f_2,…,f_i\}$. Each graph can be represented with the help of the adjacency matrix A. Thus, we implement the feature graph modeling of the binary PE file.

\begin{figure}[t]
	\centering
	\includegraphics[width=0.49\textwidth]{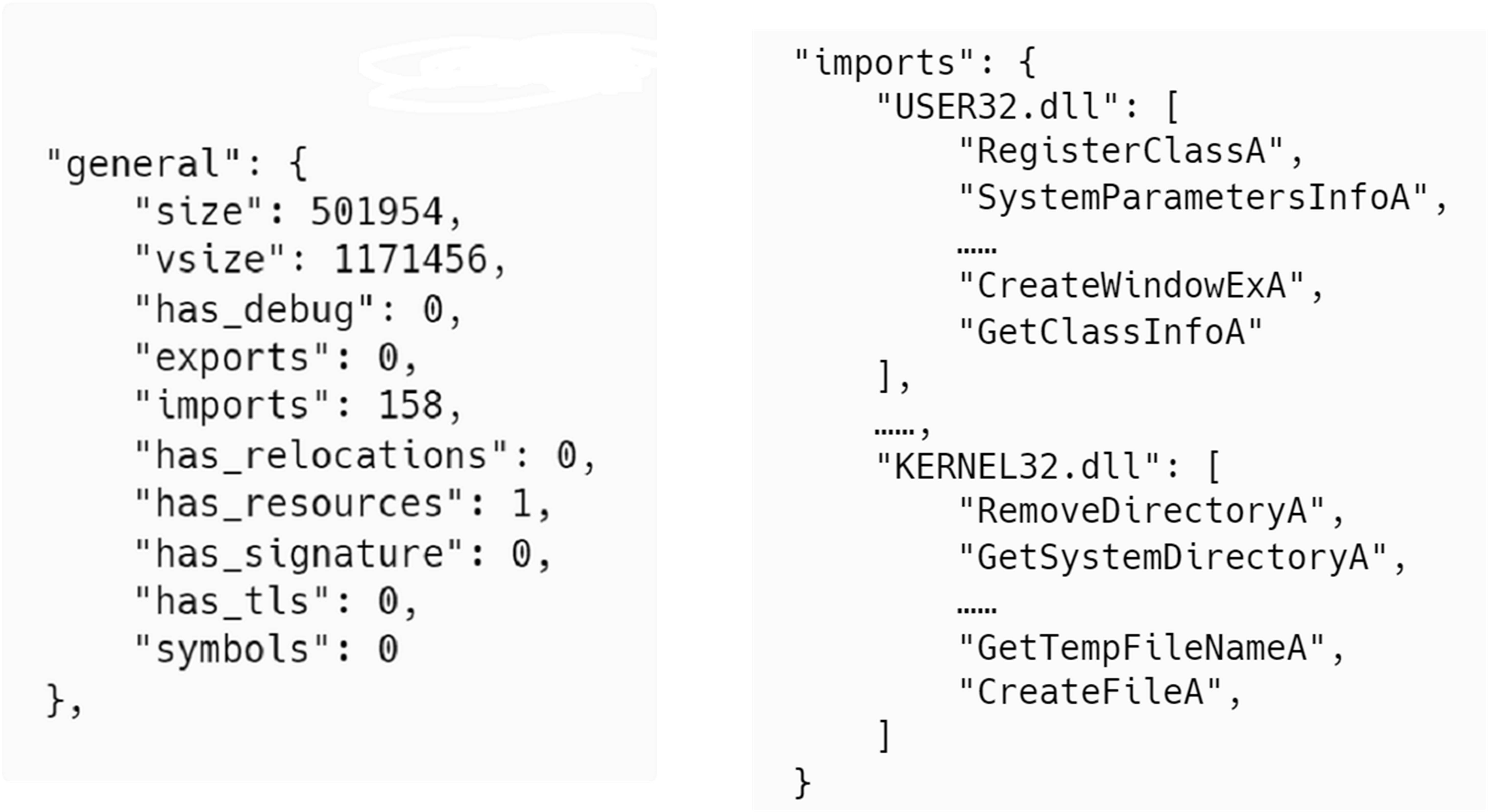}
	\caption{Static features extracted from binary PE files using LIEF.}\label{general-import}
\end{figure}

\begin{figure}[t]
	\centering
	\includegraphics[width=0.49\textwidth]{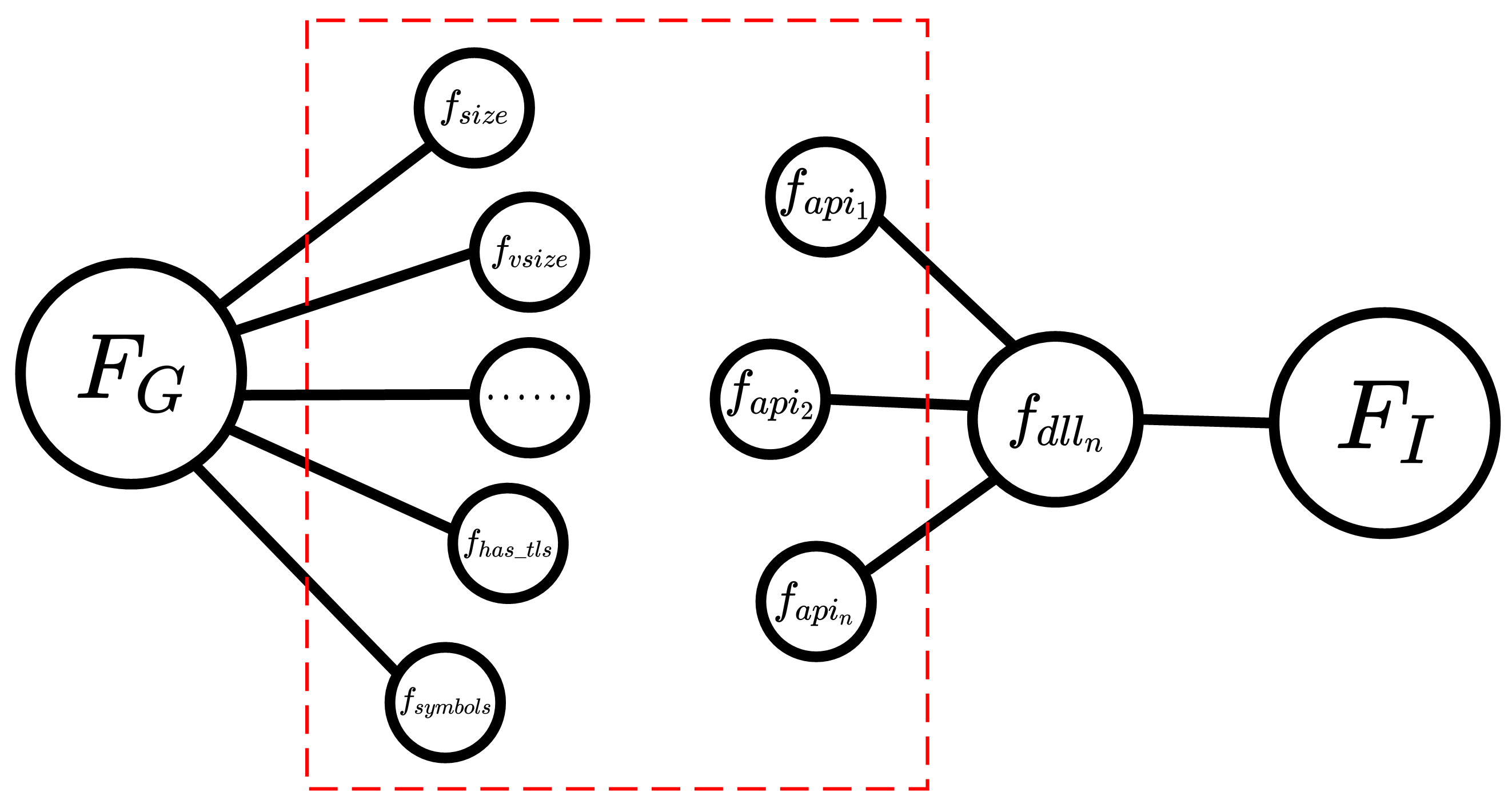}
	\caption{Construction of feature graph.}\label{general-import-graph}
\end{figure}

The specific process of FGM can be described as follows: after obtaining the static features of the binary PE file, the feature format obtained is shown in Figure \ref{general-import}. We divide the features of malware into first-level features and second-level features, and first vectorize the former features. When vectorizing the first-level features, we first obtain the number of second-level features of the first-level features. If there are third-level features of the second-level features, then the second-level feature vector is composed of third-level features, and the first-level feature vector is composed of second-level features. $F_I$, the primary feature Imported, contains multiple second-level features $f_{dll_n}$. $f_{dll_n}$ indicates the name of the exported DLL function. In addition, a significant amount of API calls will be made by the DLL, so the secondary feature $f_{dll_n}$ also contains the tertiary feature $f_{api_i}$, which is shown as two nodes in the feature graph. $F_G$ indicates general characteristics, including file size, resources, signatures and etc. As none of these secondary features have three-level features, the primary feature $F_G$, which consists of secondary features, is only shown as a node in the feature graph, and the node in the dashed box in Figure \ref{general-import-graph} will not appear in the feature graph.

\subsection{Graph representation learning}

\subsubsection{Graph convolution layer}

Inspired by Zhang et al. \cite{zhang2018end}, we use a deep graph convolutional network for graph representation learning in a malware detection task. We denote the node attribute matrix and adjacency matrix of a graph $G=\left(V,E\right)$ containing $n$ vertices as $X\in R^{n\times c}$ and $A\in Z^{n\times n}$, respectively. $G$ is an undirected graph, where $n$ is the number of vertices and each vertex is a c-dimensional feature vector. The process of graph convolution is given in the following equation:

\begin{equation}
	Z=f\left({\widetilde{D}}^{-1}\widetilde{A}XW\right)
	\label{Eq.Z}
\end{equation}

Where $\widetilde{A}=A+I$ denotes the adjacency matrix with the self-loop added to enable the propagation of the properties of the vertex itself back to itself. $\widetilde{D}$ denotes the corresponding diagonal matrix, $W\in R^{c\times c^\prime}$ denotes the trainable parameter matrix, $f$ denotes the nonlinear activation function, and $Z \in R^{n\times c^\prime}$ is the output matrix after activation.

The graph convolution operation propagates the attributes of each vertex to its neighbors. In this process, in order to aggregate the attributes at multiple scales, we perform multiple graph convolution operations and recursively perform the feature scale transformation as,

\begin{equation}
	Z^{t+1}=f\left({\widetilde{D}}^{-1}\widetilde{A}X^tW^t\right)
	\label{Eq.Z^{t+1}}
\end{equation}

Where $Z^0=X$, the $t\text{-}th$ layer indicates that the input is $Z^t\in Z^{n\times c_t}$, and the $c_t$ feature channels are mapped to $c_{t+1}$ by the parameter matrix $W^t\in R^{c_t\times c_{t+\mathbb{1}}}$. Based on the connectivity of the structure, each vertex's newly acquired feature channel is propagated to its neighboring nodes. Furthermore, owing to the addition of the self-loop, the new feature channel is also propagated back to itself, which is then multiplied by the adjacency matrix $\widetilde{A}$. This  in turn is normalized using ${\widetilde{D}}^{-1}$, so that the vertices in the graph are able to search in a breadth-first manner in the graph to pass their own properties. The $j\text{-}th$ feature channel of vertex i is treated as a linear combination of the $j\text{-}th$ feature channel of all neighboring nodes of that vertex. After h graph convolution layers, the output $Z_t$ of each layer is concatenated, yielding,

\begin{equation}
	Z^{1:h}=\left[Z^1,Z^2,\ldots\ldots,Z^h\right]
	\label{Eq.Z^{1:h}}
\end{equation}

\subsubsection{SortPooling layer}

\begin{figure*}[ht]
	\centering
	\includegraphics[width=0.99\textwidth]{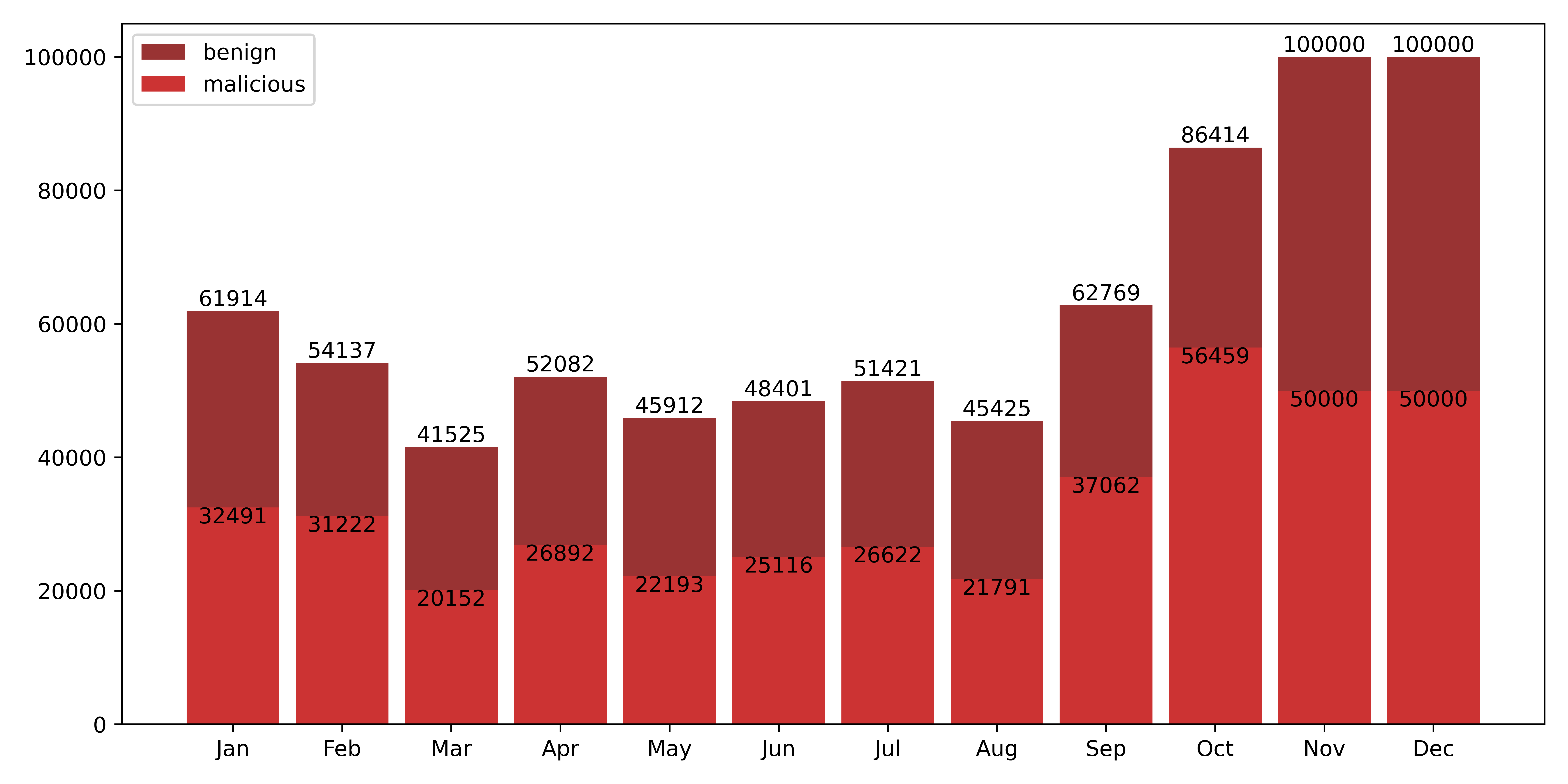}
	\caption{Distribution of samples in EMBER dataset in 2018.}\label{malware-distribution}
\end{figure*}

After multiple layers of convolution, the final output is $n$ rows and $\sum_{1}^{h}c_t$ columns of $Z^{1:h}$. Each row can be considered as a vertex "feature descriptor", which encodes the local substructure information of the node at multiple scales. $Z^{1:h}$ is approximately equal to the finest continuous color in the kernel of the Weisfeiler-Lehman graph \cite{shervashidze2011weisfeiler}, and then those are used to sort all vertices. Sorting is performed  in decreasing order according to the last channel of the last layer. If two vertices have the same value in the last channel, $Z^{h-1}$ of the last layer is used to continue sorting, and so on. In this way, it is possible to obtain a representation of the entire graph, which can then be trained using traditional neural networks. To improve model generalization and unify output, similar to the work \cite{rong2019dropedge}, a hyperparameter called pooling rate needs to be adjusted to prevent overfitting of the model by discarding some of the feature channels. The pooling rate is used to determine the number of channels $k$ that retain the final output of $Z^{1:h}$. If $n>k$, the last $n\text{-}k$ rows are deleted; if less than $k$, $k\text{-}n$ rows  need to be filled so that the output has $k$ rows, so that each graph can obtain a uniform feature output $Z^{sp}$.

\subsubsection{Remaining layer}

In the original DGCNN \cite{zhang2018end}, a tensor $Z^{sp}$ of size $k\times\sum_{1}^{h}c_t$ is output after the sorting pooling layer, with each row representing a vertex and each column representing a feature channel, and finally the authors add a one-dimensional convolutional kernel of size $\sum_{1}^{h} c_{t}$ with a step size $\sum_{1}^{h} c_{t}$ such that the output $Z^{sp}$ of the sorting pooling layer ends up as a matrix of size $k \times F$, and $F$ is the number of convolution kernels. This has been improved in the research \cite{Yan2019Classifying}. The single-channel Conv1D layer is parametrically represented as $W\in R^{\mathbb{1}\times k}$ and the output $E\in R^{\mathbb{1}\times\sum_{\mathbb{1}}^{h}c_t}$, which is implemented as shown in the following equation, where $f$ is the nonlinear activation function. 

\begin{equation}
	E=f\left(W\times Z^{sp}\right)
	\label{Eq.E}
\end{equation}

Based on the idea of graph embedding \cite{xu2017neural}, the output $Z_i^{sp}$ of each row of the sorted pooling result is used as the embedding of the retained vertices, and then the weighted sum of the vertices, which is also the embedding $E_G$ of the graph $G$.

\subsection{Classifier module}

Building on earlier work \cite{Yan2019Classifying}, we added a multilayer perceptron after the output of the graph convolution layer for malware detection task.
The multi-layer perceptron \cite{Pal1992Multilayer}, a deep learning classification model, comprises three distinct components: an input layer, an output layer, and a hidden layer, which may consist of multiple fully connected layers. In order to enhance network performance and avoid overfitting, we have incorporated a Dropout layer \cite{hinton2012improving} subsequent to every fully connected layer within the hidden layer, as,

\begin{equation}
	H=\sigma\left(E_GW_h+b\right)
	\label{Eq.H}
\end{equation}

Where the weight parameters and deviation parameters of the hidden layer are denoted by $W_h$ and $b$, respectively. $\sigma$ denotes the activation function. We used softmax as the activation function to transform the output values of the classification into a probability distribution between $\left[0,1\right]$.

\section{Experiments}
\label{Experiments}

\subsection{Experiment setup}

\subsubsection{Dataset}

The EMBER dataset, a considerably large benchmark dataset of malware, has been published by Endgame. EMBER comprises features that have been extracted from 1.1M binary samples, where 900K samples are allotted for training purposes, consisting of 300K malicious, benign, and untagged samples each. Moreover, there are an additional 200K samples reserved for testing, which includes 100K malicious as well as benign samples respectively. In order to analyze the impact of concept drift, we have systematically categorized 800K samples based on a timeline. Figure \ref{malware-distribution} illustrates the distribution of samples from January to December 2018 in the EMBER dataset and provides specific values. As the samples prior to 2018 contained only malicious samples and no benign samples, they were not used for training and were therefore discarded.

\subsubsection{Evaluation metrics}

\begin{figure}[t]
	\centering
	\includegraphics[width=0.49\textwidth]{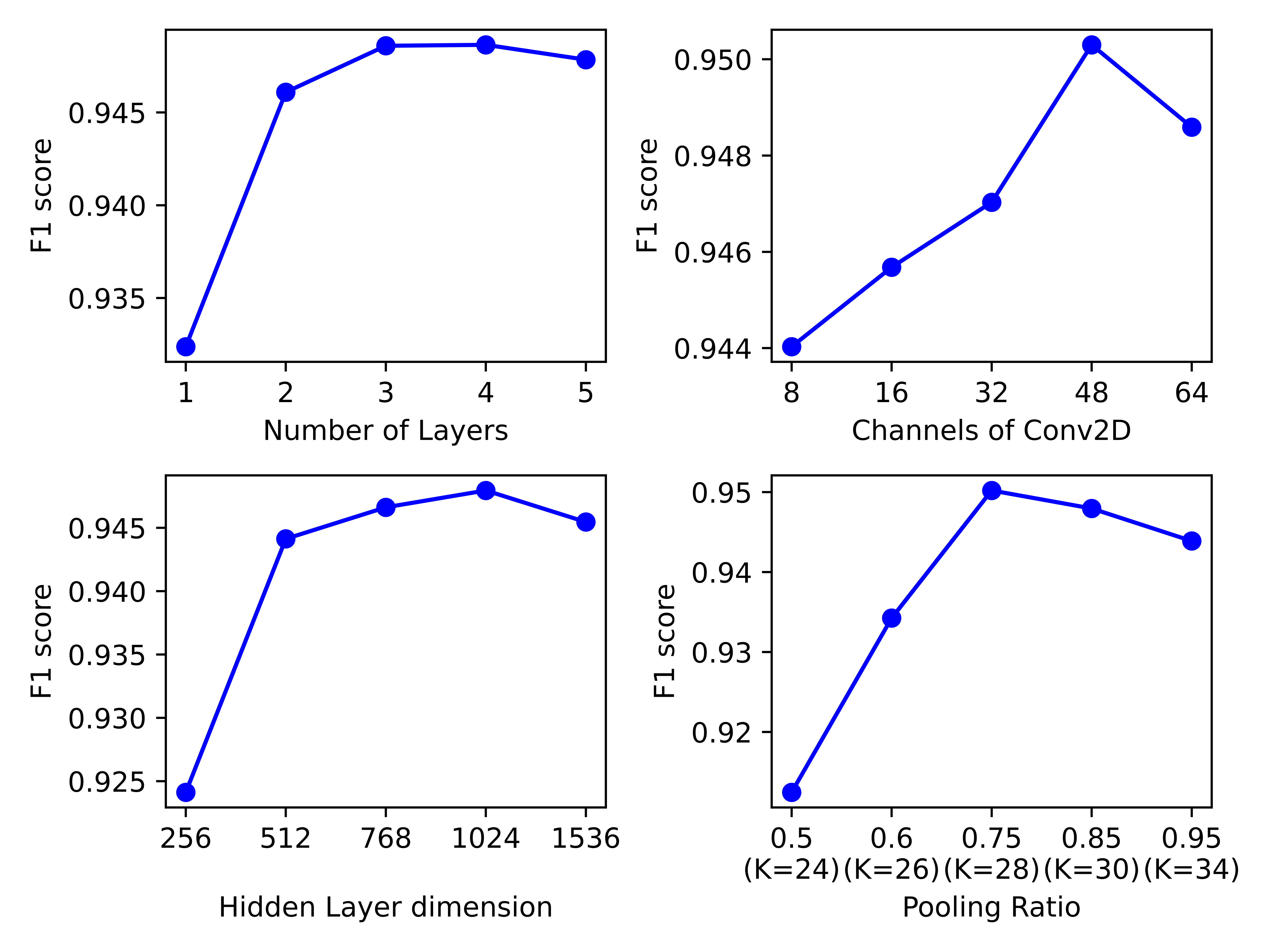}
	\caption{Hyper parameters optimization.}\label{hyper_parameters}
\end{figure}

In the malware detection task, as in Ref. \cite{Wang2019Heterogeneous}, we use AUC score, accuracy, and F1 score to evaluate the performance of various methods. The accuracy and F1 score are calculated as in the following equations:

\begin{equation}
	Accuracy=\frac{TP+TN}{TP+TN+FP+FN}
	\label{Eq.Accuracy}
\end{equation}

\begin{equation}
	F1-score=\frac{2\times Precision \times Recalll}{Precision + Recall}
	\label{Eq.F1-score}
\end{equation}

Where $Precision=\frac{TP}{TP+FP}$, $Recall=\frac{TP}{TP+FN}$

\subsection{Hyper parameter selection}

\begin{table*}[t]\centering
	\caption{Performance on malware detection.}
	\label{Performance on Malware Detection}
	\begin{tabular}{c|cccc}
		\hline
		\textbf{Methods}  &\textbf{Dataset(date)}  &\textbf{AUC score}  &\textbf{Accuracy}  &\textbf{F1 score}\\
		\hline
		LR-based \cite{Roopak2020TAN}														&01/2018	&0.74687	&0.67603	&0.67322	\\
		DT-based \cite{Abel2019Malware}													&01/2018	&0.91433	&0.91513	&0.91481	\\
		SVM-based \cite{Enhanced2020Hyoil,Wadkar2020Detecting}										&01/2018	&0.57845	&0.53060	&0.43027	\\
		KNN-based \cite{Kong2022PMMSA}		   &01/2018	   &0.95048	   &0.89599	   &0.89585	\\
		AdaBoost-based \cite{Wei2021Lightweight} 								&01/2018	&0.93596	&0.86030	&0.85999	\\
		MLP-based \cite{Wei2021Lightweight,Qiao2021Malware}		  									 &01/2018	 &0.67127	  &0.68621	   &0.64664	\\
		FFDNN \cite{singh2023feed}		  									 &01/2018	 &0.70460	  &0.61176	   &0.69156	\\
		MFGraph (ours)																														&01/2018	&0.98756	&0.95026	&0.95020	\\
		\hline
	\end{tabular}
\end{table*}

To search for the hyperparameters of MFGraph, we employed five-fold cross-validation. We used 80\% of the data from January 2018 for training, which was divided equally into five copies. In each non-repeating cycle, we used four copies (80\%) for training and the remaining one (20\%) for validation. During the experiment, we trained each model for 20 epochs and recorded the F1 score of each epoch to determine the optimal model. MFGraph has four main hyperparameters to search, namely, pooling rate, number of layers, number of channels, and number of hidden neurons. We plotted the results of these hyperparameters in Figure \ref{hyper_parameters}, where the y-axis represents the F1 score, and the x-axis represents the search values of the hyperparameters.

Our findings demonstrate that MFGraph attains its highest F1 score when configured with three layers, 1024 neurons, 48 convolutional channels, and a pooling rate of 0.75 (K=28). While larger hyperparameters may consume more computational resources, they have little impact on F1 score improvement and may even result in degraded performance. As such, we have incorporated this optimal set of hyperparameters into the default parameters of MFGraph for malware detection tasks.

\begin{figure}[t]
	\centering
	\includegraphics[width=0.49\textwidth]{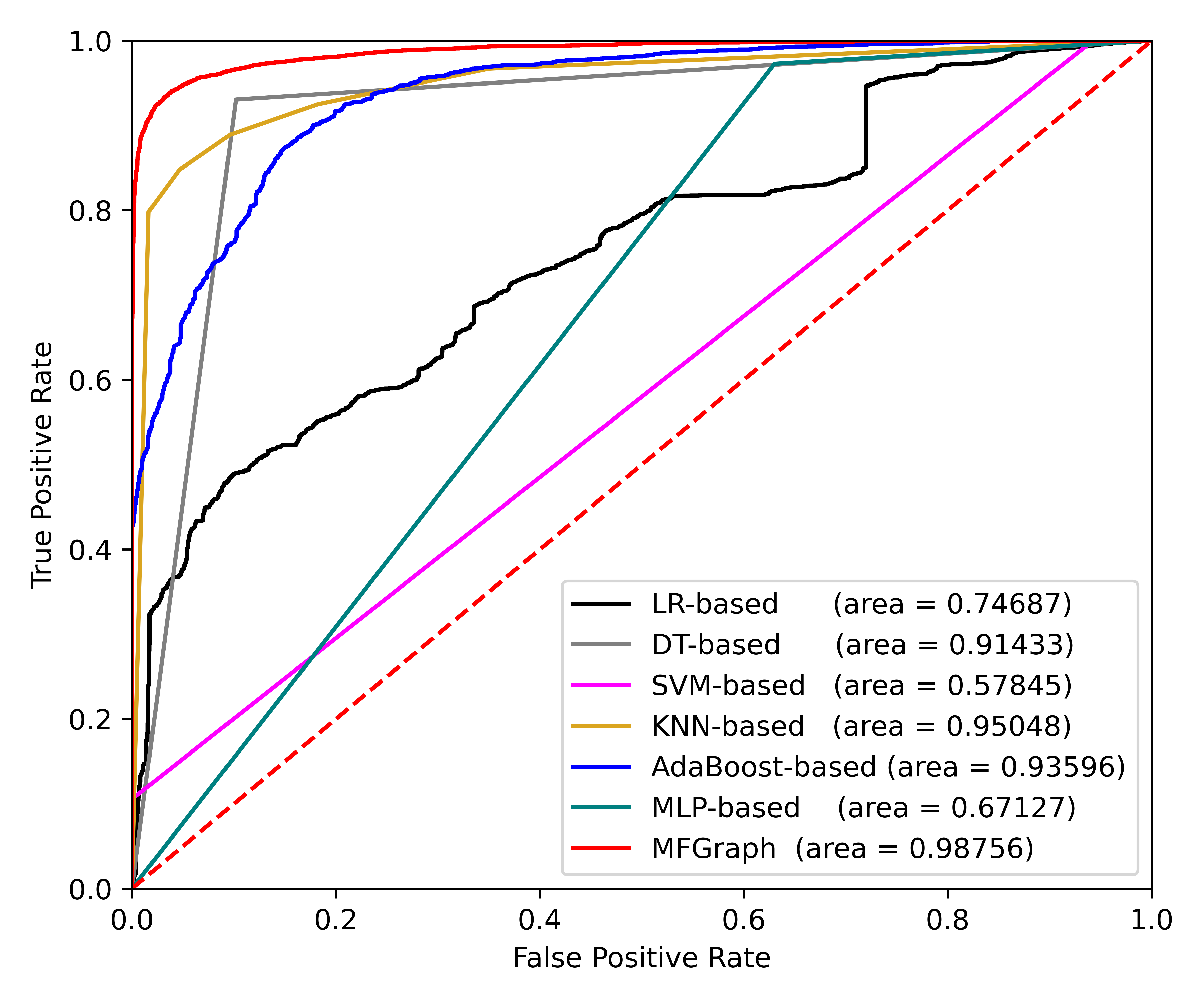}
	\caption{ROC curves on detecting malware.}\label{ROC_AUC_score}
\end{figure}

\begin{table*}[t]\centering
	\caption{AUC score comparison of MFGraph and other methods on 12 test subsets.}
	\label{AUC score changes}
	\begin{tabular}{c ccccccc}
		\hline
		\multirow{2}{*}{\textbf{Methods}} & 
		\multicolumn{7}{c}{\textbf{AUC score}}\\
		\cline{2-8} 
		&\begin{tabular}[c]{@{}c@{}} LR-based\\ \cite{Roopak2020TAN}\end{tabular}			&\begin{tabular}[c]{@{}c@{}} DT-based\\ \cite{Abel2019Malware}\end{tabular}			&\begin{tabular}[c]{@{}c@{}} SVM-based\\ \cite{Enhanced2020Hyoil,Wadkar2020Detecting}\end{tabular}	&\begin{tabular}[c]{@{}c@{}} KNN-based\\ \cite{Kong2022PMMSA}\end{tabular}		&\begin{tabular}[c]{@{}c@{}} AdaBoost-based\\ \cite{Wei2021Lightweight} \end{tabular}	&\begin{tabular}[c]{@{}c@{}} MLP-based\\ \cite{Wei2021Lightweight,Qiao2021Malware}	\end{tabular}	&\begin{tabular}[c]{@{}c@{}} MFGraph\\ (ours)\end{tabular}	\\
		\hline
		Test-02/2018 	&0.74631	&0.82830	&0.53633	&0.80225	&0.89842	&0.63280	&0.96145	\\
		Test-03/2018 	&0.71211	&0.82966	&0.53177	&0.77295	&0.90459	&0.66013	&0.95054	\\
		Test-04/2018 	&0.71198	&0.87515	&0.51877	&0.72141	&0.92780	&0.69681	&0.97117	\\
		Test-05/2018 	&0.64027	&0.81765	&0.51693	&0.73212	&0.92948	&0.70331	&0.96239	\\
		Test-06/2018 	&0.67793	&0.80858	&0.51073	&0.70680	&0.89345	&0.65006	&0.94826	\\
		Test-07/2018		&0.62644	&0.83340	&0.51284	&0.71804	&0.91473	&0.70509	&0.96436	\\
		Test-08/2018		&0.66877	&0.80484	&0.50307	&0.68703	&0.91463	&0.67119	&0.95061	\\
		Test-09/2018		&0.71910	&0.79059	&0.50390	&0.70513	&0.86831	&0.65258	&0.94210	\\
		Test-10/2018		&0.67987	&0.75276	&0.51105	&0.65259	&0.78686	&0.69077	&0.92989	\\
		Test-11/2018		&0.71343	&0.75666	&0.51888	&0.67723	&0.80486	&0.65839	&0.94277	\\
		Test-12/2018		&0.70543	&0.74691	&0.51229	&0.69278	&0.78674	&0.62879	&0.92872	\\
		\hline
	\end{tabular}
\end{table*}

\begin{figure*}[t]
	\centering
	\includegraphics[width=0.99\textwidth]{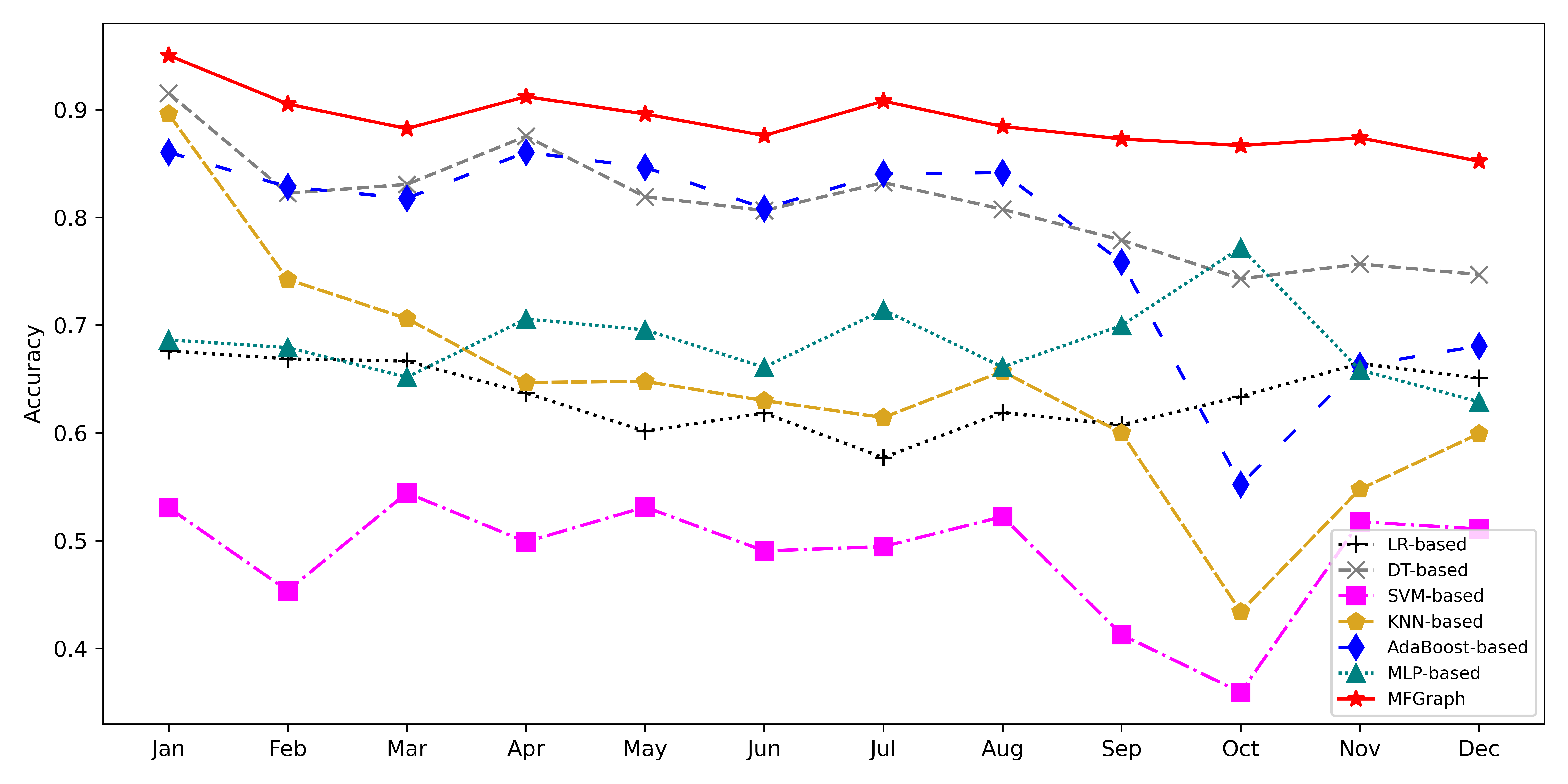}
	\caption{Variations in accuracy for MFGraph versus other methods on 12 test subsets.}
	\label{accuracy}
\end{figure*}

\subsection{Malware detection}

The first concern we studied is the accuracy of MFGraph in detecting malware. According to the distribution of EMBER dataset, the training set was composed of 80\% of all samples collected in January 2018, while the remaining 20\% were utilized as the test set. Upon completing 20 epochs, the model converged and demonstrated a high level of performance, with an AUC score of 0.98756, an accuracy of 0.95026, and an F1 score of 0.95020. The accompanying ROC curve can be found in Figure \ref{ROC_AUC_score}. In addition, Table \ref{Performance on Malware Detection} provides a comparison of the model's results to those of six other detection methods on the test set. MFGraph outperforms six other methods according to experimental results, particularly in terms of AUC score, MFGraph outperforms all of the baseline models by at least 3.708\%, with differences ranging from 4.0911\% to 31.629\%. Specifically, the AUC score of MFGraph is 24.069\% and 7.323\% higher than that of LR-based, and DT-based models. Additionally, both accuracy and F1 scores of MFGraph are at least 3.513\% and 3.539\% higher than those of the other models. The results indicate that our proposed MFGraph is capable of performing well in malware detection even when dealing with an unbalanced distribution in the dataset. This experimental finding provides evidence of the effectiveness of MFGraph in malware detection, as it outperforms other models such as DT-based, SVM-based, KNN-based, AdaBoost-based, and MLP-based in terms of AUC score, accuracy, and F1 score.

FFDNN \cite{singh2023feed} is a neural network model with a deeper structure compared to MLP-based models, featuring more neurons and network layers. However, the number of neurons and layers does not necessarily correlate with improved performance. As shown in Table  \ref{Performance on Malware Detection}'s experimental results, the shallower MLP model achieved higher accuracy, while FFDNN scored better on AUC and F1 metrics. There are notable performance differences in malware detection between SVM-based and KNN-based methods. Ref. \cite{singh2022investigation} and Ref. \cite{singh2022minimized} delve into these variances, emphasizing the impact of data preprocessing on performance. The latter method combines feature selection techniques to preserve maximum relevant information with minimal redundancy. Furthermore, Ref. \cite{singh2022performance} preprocesses PE file header information of malware using various methods and employs an SVM model for detection. While these approaches reduce some feature overheads, they still necessitate laborious feature selection processes. In contrast, MFGraph only requires extracting nine static features through the LIEF parser and organizing them into a feature graph. By utilizing a multilayer perceptron, it achieves precise detection while offering interpretability regarding connections among different features.

The superior performance of MFGraph over all other baselines can be attributed to its implementation of graph structures. This is because such structures encapsulate more semantic information and connections between the various feature nodes, leading to a more comprehensive understanding of the underlying knowledge. In contrast, many previous methods generally utilize direct concatenation of features without considering any intrinsic relationships between them, resulting in suboptimal malware characterization and diminished detection capabilities. Direct concatenation of features in MLP-based detection yields an accuracy  of only 0.68621, whereas the accuracy of learned representation vectors improves by 26.405\% to 0.95026 when classifying them using MLP after first modeling the features with graph structures. Notably, our method outperforms all others in terms of AUC score, accuracy, and F1 score.

\subsection{Impact of concept drift}

\begin{figure*}[t]
	\centering
	\includegraphics[width=0.99\textwidth]{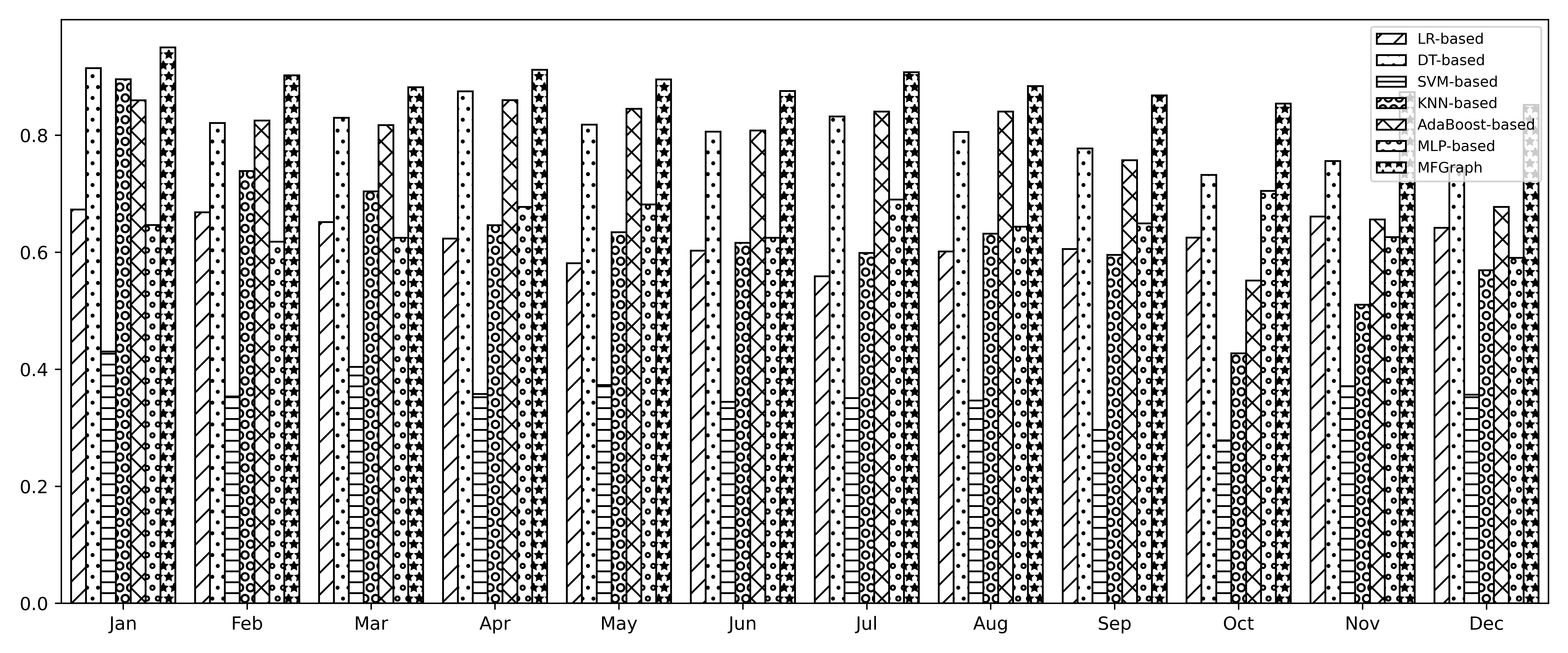}
	\caption{Comparing F1 score variations of MFGraph to other methods on 12 test subsets.}
	\label{F1}
\end{figure*}

To evaluate the effect of concept drift, we employed the classifier that was trained on the January 2018 dataset to classify a new dataset consisting of data from February through December 2018. The dataset was partitioned into 12 subsets, with each subset representing a month of the year based on the time series. In Figure \ref{malware-distribution}, we present the distribution of benign and malware in the EMBER dataset for each month of the year 2018, accompanied by their respective quantitative values. To predict data from February to December, we employed models trained on data collected during January 2018 and evaluated their performance using the AUC score, accuracy, and F1 score metrics. Table \ref{AUC score changes} presents the AUC score for MFGraph in comparison to other methods, while Figure \ref{accuracy} displays the accuracy trend plots over time for different methods. Additionally, Figure \ref{F1} provides F1 scores for each method across 12 test subsets.

The experimental results indicate that previous detection methods, including LR-based, DT-based, SVM-based, KNN-based, AdaBoost-based, and MLP-based approaches, exhibit significant degradation in performance when applied to new data due to changes in data distribution or the emergence of new malware variants. This is particularly evident when classifiers trained on older datasets are utilized to classify new data. In the initial months, most detection methods demonstrate a relatively stable performance. However, their efficacy dwindles over time due to the presence of concept drift in the actual environment. Known data trained models face difficulty in predicting unknown malware variants because new variants often have some features added or removed. Existing methods typically concatenate features without considering their inter-relationships, which restricts access to global features of malware. Therefore, when concept drift occurs, models undergo degradation, leading to poor performance. The detection method proposed in this study, known as MFGraph, incorporates potential relationships between features by utilizing graph representation learning methods to perform feature embedding. This process results in a set of representation vectors possessing global properties, leading to enhanced stability in the detection performance. 

\begin{figure*}[t]
	\centering
	\includegraphics[width=0.99\textwidth]{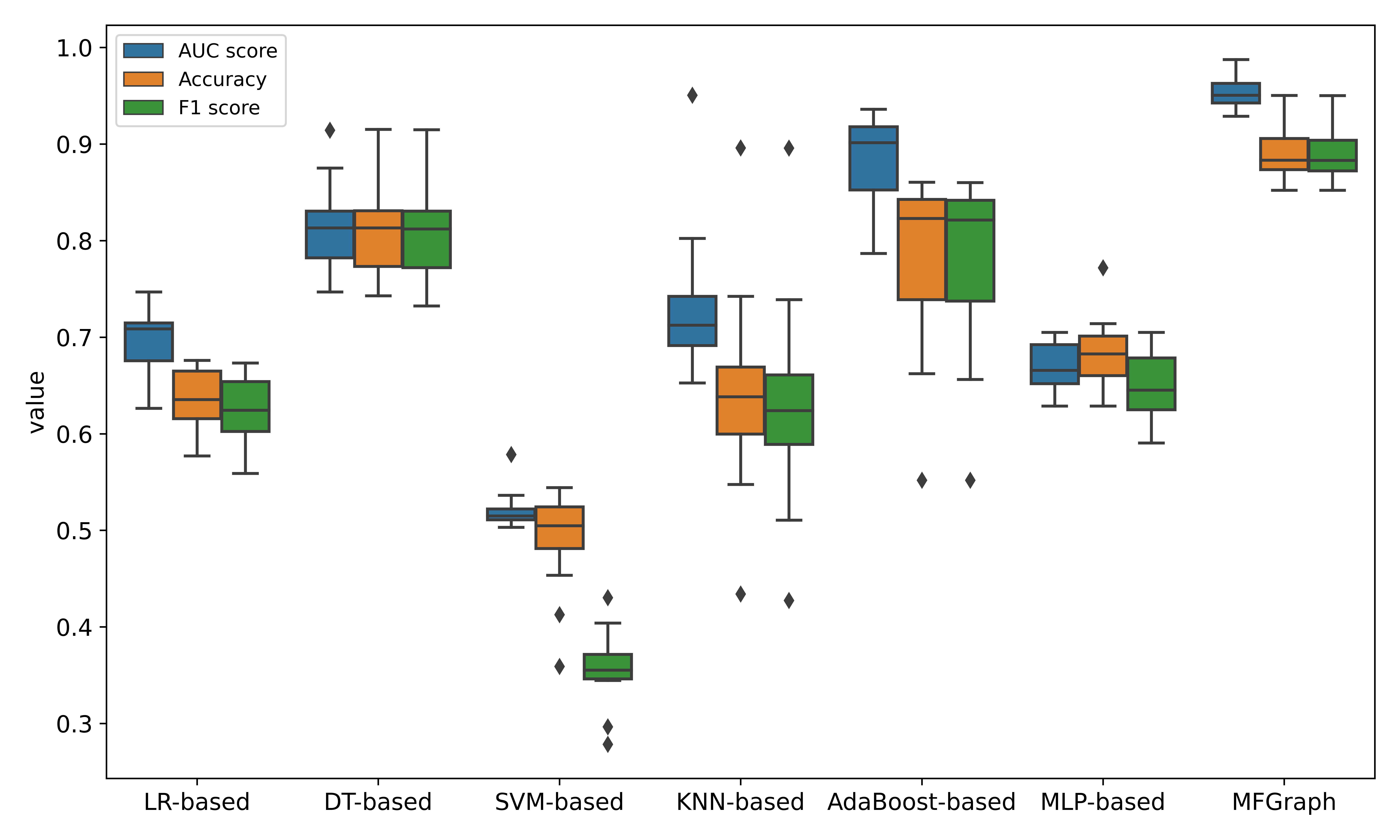}
	\caption{Performance changes of different malware detection methods under concept drift on 12 test subsets.}
	\label{box}
\end{figure*}

To comprehensively assess the impact of concept drift, we not only record the highest and lowest AUC score, accuracy, and F1 score achieved by each model tested on all 12 subsets in Table \ref{changes} but also calculate the value of the change in each model's three metrics, specifically the maximum percentage point of decline. Additionally, the visual representation of this change is illustrated in Figure \ref{box}. The experimental results reveal that among the six models tested, KNN-based is the most susceptible to the impact of concept drift, with a significant AUC drop of 29.789\% and corresponding drops in both accuracy and F1 score by as much as 46.193\% and 46.853\%, respectively, indicating the most notable degradation. In contrast, our proposed MFGraph demonstrates superior performance. Specifically, within the first six months, MFGraph maintains consistently high AUC, with a decrease of less than 4\%, while other models, including LR-based, DT-based, SVM-based, KNN-based, AdaBoost-based, and MLP-based, experience decreases ranging from 6.575\% to 24.368\%.  In the evaluation of all tested subsets, our proposed MFGraph displays superior performance with the highest AUC score, accuracy, and F1 score. Furthermore, compared to the maximum value, MFGraph exhibits only a marginal decrease in the AUC score by 5.884\%, indicating its robustness to changes in the dataset. In contrast, although SVM-based shows the least change in the AUC score among the other six models, it performs poorly overall, with a maximum AUC score of only 0.57845. Similarly, while LR-based has the least change in accuracy and F1 scores, its maximum values for these metrics are much lower than the performance achieved by MFGraph, with a maximum accuracy of 0.67603 and a maximum F1 score of 0.67322. The  results demonstrate that our proposed MFGraph is highly effective in mitigating the impact of concept drift, evident from its superior performance and stability compared to other existing malware detection methods.

\begin{table*}[t]\centering
	\caption{Evaluation of best and worst AUC score, accuracy, and F1 score on test subsets, along with percentage of performance degradation. DegRate denotes degradation rate.}
	\label{changes}
	\resizebox{\textwidth}{!}{
		\begin{tabular}{c|ccc|ccc|ccc}
			\hline
			\multirow{2}{*}{\textbf{Methods}} & \multicolumn{3}{c|}{\textbf{AUC score}} & \multicolumn{3}{c|}{\textbf{Accuracy}} & \multicolumn{3}{c}{\textbf{F1 score}}\\
			\cline{2-10}
			&\textbf{Best}	&\textbf{Worst}		&\textbf{DegRate}		&\textbf{Best}	&\textbf{Worst}		&\textbf{DegRate}		&\textbf{Best}	&\textbf{Worst} 	&\textbf{DegRate}		\\
			\hline
			LR-based\cite{Roopak2020TAN} 			&74.687\%	&62.644\%	&12.043\%	&67.603\%	&57.702\%	&9. 901\%	&67.322\%	&55.894\%	&11. 428\%	\\
			DT-based\cite{Abel2019Malware} 			&91.433\%	&74.691\%	&16.742\%	&91.513\%	&74.295\%	&17.218\%	&91.481\%	&73.236\%	&18. 245\%	\\
			SVM-based\cite{Enhanced2020Hyoil,Wadkar2020Detecting} 			&57.845\%	&50.307\%	&7.538\%	&54.432\%	&35.911\%	&18.521\%	&43.027\%	&27.854\%	&15.173\%	\\
			KNN-based\cite{Kong2022PMMSA} 			&95.048\%	&65.259\%	&29.789\%	&89.599\%	&43.406\%	&46.193\%	&89.585\%	&42.732\%	&46. 853\%	\\
			AdaBoost-based\cite{Wei2021Lightweight} 	&93.596\%	&78.674\%	&14.922\%	&86.041\%	&55.197\%	&30.844\%	&86.005\%	&55.186\%	&30.819\%	\\
			MLP-based\cite{Wei2021Lightweight,Qiao2021Malware} 			&70.509\%	&62.879\%	&7.63\%		&77.183\%	&62.878\%	&14.305\%	&70.492\%	&59.054\%	&11.438\%	\\
			MFGraph(ours) 		&98.756\%	&92.872\%	&5.884\%	&95.026\%	&85.207\%	&9.819\%	&95.020\%	&85.206\%	&9.814\%	\\
			
			\hline
		\end{tabular}
	}
\end{table*}

\section{Discussion}
\label{Discussion}

\subsection{Construction forms of MFGraph}

\begin{figure*}[htbp]
	\centering
	\includegraphics[width=0.99\textwidth]{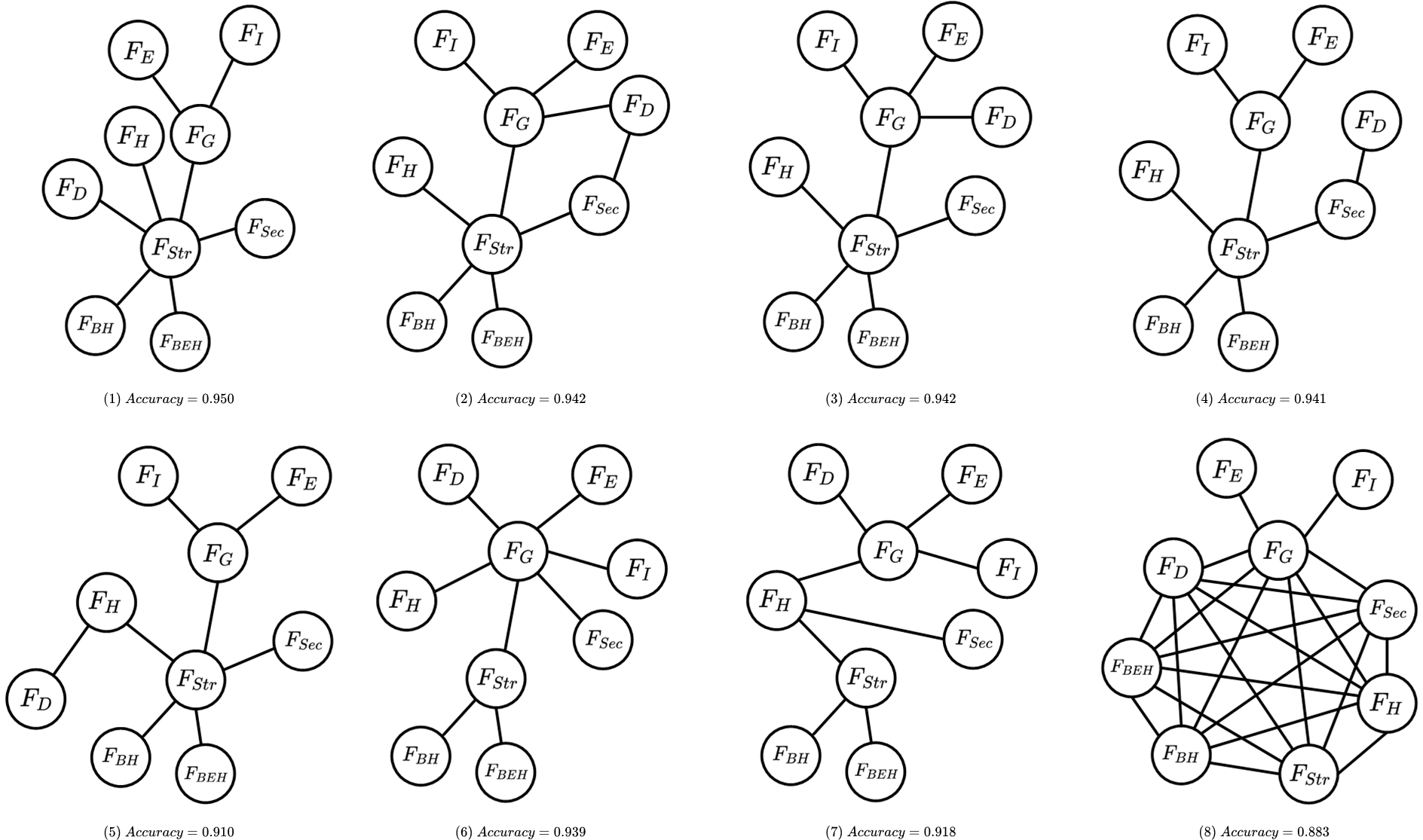}
	\caption{Different construction forms of nine first-level features $F_G$, $F_H$, $F_I$, $F_E$, $F_{Sec}$, $F_{BH}$, $F_{BEH}$, $F_{Str}$, $F_D$ and corresponding accuracy.}
	\label{GRAPHS}
\end{figure*}

We outlined the step-by-step construction of the feature graph in Section \ref{mfgraph_con}. In graph representation learning, the broadcast mechanism enables non-adjacent nodes to access each other's information, creating numerous ways to link nine static features. To investigate alternative methods for constructing the feature graph, we conducted a series of experiments. Figure \ref{GRAPHS}(2) illustrates that connecting $F_D$ as an intermediary node between $F_G$ and $F_{Sec}$ without linking to $F_{Str}$ resulted in a decrease in overall detection accuracy to 0.942, suggesting that $F_D$ contains more relevant data directory information from $F_{Str}$. Removing the connection between $F_D$ and $F_{Sec}$ (Figure \ref{GRAPHS}(3)) did not impact accuracy, indicating their linkage is not crucial. However, eliminating the connection between $F_G$ and $F_{Str}$ (Figure \ref{GRAPHS}(4)) reduced accuracy to 0.941, revealing a weak relationship between $F_D$ and $F_G$. Hence avoiding direct connections with performance representation may enhance capturing characteristics from both sides ($F_D$ and $F_{Sec}$).

Due to the valuable information in $F_D$, including resource directory, relocation directory, debug directory, and loading configuration directory, these features can complement $F_{Str}$ for a more comprehensive global representation performance. We initially linked $F_D$ to $F_H$ as illustrated in Figure \ref{GRAPHS}(5). However, experimental results show that this linking method notably decreases detection performance. Consequently, we also connected $F_D$ to $F_G$ as shown in Figure \ref{GRAPHS}(6). In contrast to Figure \ref{GRAPHS}(3), where $F_G$ is additionally connected with $F_H$ and $F_{Sec}$ but only achieves an accuracy of 0.939. Moreover, as depicted in Figure \ref{GRAPHS}(7), we tried using $F_H$ as an intermediary node by connecting it with both $F_{Str}$ and $F_G$ and associating it with $F_{Sec}$. Nevertheless, the accuracy was only 0.918. This highlights a direct correlation between $F_G$ and $F_{Str}$, where utilizing $F_H$ as an intermediary node for propagation is unsuitable. It is important to note that creating a circular structure by linking pairs of nodes ($F_G$, $F_{Sec}$, $F_H$, $F_{Str}$, $F_{BH}$, $F_{BEH}$, $F_D$) while each node is also connected with four other non-adjacent nodes resulted in the lowest accuracy shown in Figure \ref{GRAPHS}(8). This confirms that certain features do not require direct connections. Otherwise, there will be significant redundant information during graph representation learning leading to a noticeable decline in malware detection performance.

\subsection{Complexity and interpretability}

MFGraph utilizes graph structure to model malware's static characteristics, conducts graph representation learning via deep graph convolutional neural networks, and ultimately performs malware detection using a three-layer perceptron classifier. Deep graph convolutional networks are well-suited for handling data with graph structures, effectively capturing intricate dependencies between nodes. However, this increases the model's complexity. Table \ref{Performance on Malware Detection} demonstrates that employing a perceptron-based approach directly to learn classification sample features results in a low accuracy of only 0.68 due to the loss of valuable information between features encoded as data. By initially encoding features using a graph structure and leveraging deep graph convolutional networks to learn potential relationships among features before classifying feature vectors rich in information through perceptron, the malware detection performance significantly improves to 0.95. This highlights that graph convolution offers better interpretability and can identify complex dependencies among different features, simple concatenation would lead to substantial information loss. Furthermore, Figure \ref{GRAPHS} reveals that most static malware is associated with $F_{Str}$, the relationship between $F_D$ and $F_{Sec}$ is insignificant compared to that between $F_D$ and $F_{Str}$ which has a more significant impact on detection outcomes than those between $F_D$ and $F_G$ or $F_D$ and $F_{Sec}$. Thanks to graph convolutional networks, MFGraph achieves superior performance while also providing better interpretability compared to alternative methods.

Regarding the three-layer perceptron classifier in MFGraph, it is evident that the more layers the network has, the higher the model complexity. Figure \ref{hyper_parameters} illustrates from experimental result that F1 scores for three and four-layer perceptron are nearly identical when transitioning from one to five layers. However, adding more layers can lead to decreased performance, possibly due to overfitting. Fewer layers may limit the network's non-linear expressive capacity and suboptimal detection performance. Consequently, we opted for a three-layer perceptron as MFGraph's final classifier. The experiments and analysis reveal a significant rise in network complexity after incorporating graph convolutional networks and three-layer perceptron in MFGraph. This enhancement has boosted accuracy by 27\%, improving its ability to cope with concept drift while also enhancing interpretability somewhat—a trade-off we consider valuable.

\section{Conclusion and future work}
\label{Conclusion and future work}

In this research, a feature graph-based method MFGraph is proposed to detect malware. The binary PE files are conceptualized as a feature graph comprising distinct feature nodes, thereby facilitating the learning of correlations between these features. Subsequently, a deep graph convolutional network is employed to derive the embedding of the entire graph based on the constructed feature graph. Finally, the resulting learned representation vectors are inputted into a classifier composed of MLPs to generate the output results. According to the experimental results, MFGraph not only surpasses other linear concatenation-based baseline methods in AUC score, accuracy, and F1 score but also demonstrates its ability to effectively mitigate the effects of concept drift while exhibiting better stability.

Due to the limitation of the dataset, we are currently unable to evaluate the method's performance changes over extended periods in the presence of concept drift. Meanwhile, the constructed feature graph in this research does not consider in heterogeneity among feature nodes. In upcoming studies, we will conduct a thorough analysis of the impact of various feature types on characterization performance. Additionally, this paper does not evaluate the method's performance on the malware family classification task. Therefore, we plan to assess its ability to identify similar malware families in future studies.

\section*{Authors contributions}
\textbf{Binghui Zou}: Conceptualization; Formal analysis; Investigation; Methodology; Data curation; Validation; Visualization. Writing – original draft; Writing – review \& editing. \textbf{Chunjie Cao}: Funding acquisition; Project administration; Supervision. \textbf{Longjuan Wang}: Funding acquisition; Project administration. \textbf{Yinan Cheng}: Validation; Visualization. \textbf{Chenxi Dang}: Reformatting; Data analysis; Visualization. \textbf{Ying Liu}: Reformatting; Literature review; Visualization. \textbf{Jingzhang Sun}: Writing – review \& editing.

\section*{Acknowledgments}
This research was supported by the National Natural Science Foundation of China Enterprise Innovation and Development Joint Fund (No.U19B2044), the National Key Research and Development Program of China (No.2021YFB2700600), and the Natural Science Foundation of Hainan Province (No.621MS017).

\section*{Conflicts of interest}
The authors declare that they have no competing interests.

\section*{Data availability statement}
The data used to support the findings of this study have been deposited in the ember repository (https://github.com/elastic/ember).

\bibliographystyle{vancouver}
\bibliography{bibliography}

\end{document}